\documentclass{article}
\usepackage{spconf,amsmath,graphicx,hyperref}
\usepackage{url}
\usepackage{hyperref} 
\usepackage{bbding} 
\usepackage{multirow}
\usepackage{algorithm}
\usepackage{algorithmic}
\usepackage{colortbl}
\usepackage{cite,xcolor}
\usepackage{diagbox}
\usepackage{booktabs}
\usepackage{amsfonts}
\usepackage{amsmath} 
\usepackage{colortbl, hhline, xcolor}

\usepackage{pifont}

\title{Xi+: Uncertainty Supervision for Robust Speaker Embedding}
\name{Junjie Li$^1$, Kong Aik Lee$^{1,\dagger}$, Duc-Tuan Truong$^2$, Tianchi Liu$^3$, Man-Wai Mak$^1$ \thanks{$\dagger$: Corresponding author }}
\address{$^1$ Department of Electrical and Electronic Engineering,\\ The Hong Kong Polytechnic University, Hong Kong SAR \\
$^2$ Nanyang Technological University, Singapore 
$^3$ National University of Singapore, Singapore\\
}

\begin{document}
\ninept
\maketitle
\begin{abstract}
\vspace{-1mm}
There are various factors that can influence the performance of speaker recognition systems, such as emotion, language and other speaker-related or context-related variations.  Since individual speech frames do not contribute equally to the utterance-level representation, it is essential to estimate the importance or reliability of each frame. The xi-vector model addresses this by assigning different weights to frames based on uncertainty estimation. However, its uncertainty estimation model is implicitly trained through classification loss alone and does not consider the temporal relationships between frames, which may lead to suboptimal supervision.
In this paper, we propose an improved architecture, xi+. Compared to xi-vector, xi+ incorporates a temporal attention module to capture frame-level uncertainty in a context-aware manner. In addition, we introduce a novel loss function, Stochastic Variance Loss, which explicitly supervises the learning of uncertainty.  Results demonstrate consistent performance improvements of about 10\%  on the VoxCeleb1-O set and 11\% on the NIST SRE 2024 evaluation set.

\end{abstract}


\vspace{-1mm}
\section{Introduction}
\vspace{-2mm}

Automatic speaker recognition aims to determine a speaker’s identity solely based on their voice using machine learning algorithms \cite{lee2020two}. For each speech utterance, a front-end module extracts a fixed-dimensional speaker embedding that encapsulates speaker-discriminative information while suppressing irrelevant variability such as phonetic content and background noise \cite{snyder2018x,desplanques20_interspeech,wang2023wespeaker,10889980,yakovlev2024reshape}. Speaker recognition is then performed by computing the similarity between the embeddings of the enrollment and test utterances. 

Unlike text or images, speech signals are continuous-valued and variable in length.  Moreover, speech conveys not only speaker identity but also a wide range of variations that can degrade the quality of the extracted speaker embeddings. These include linguistic factors (e.g., language), acoustic conditions (e.g., recording channel), and speaker-intrinsic traits (e.g., emotion) \cite{wang2024overview,bai2021speaker}. As a result, different temporal frames should contribute unequally to the final fixed-dimensional speaker embedding.


The ability to handle uncertainty has been extensively investigated in prior research. In the field of computer vision, incorporating uncertainty estimation enables models to assess the confidence of input data, allowing decisions to rely more heavily on the clearer or more reliable parts of an image \cite{shi2019probabilistic, ji2023map, chang2020data}. In traditional speaker recognition systems, uncertainty modeling is also fundamental. For example, the Universal Background Model (UBM) \cite{reynolds2000speaker} and the i-vector framework \cite{dehak2010front} model speaker variability using covariance matrices associated with its Gaussian components. More recently, Lee et al. \cite{lee2021xi,liu2023disentangling} proposed to incorporate uncertainty estimation into neural speaker embedding by modeling frame-level variability through a precision matrix in  pooling layer. The so-called xi-vector approach provides a novel perspective for interpreting variability in speech signals and improves robustness in speaker representation learning.

Although the xi-vector architecture demonstrates good performance, it still exhibits several limitations. First, it estimates the frame-level variance (i.e., the precision matrix) using only two learnable linear layers with several activation functions, which may lack sufficient modeling capacity. Moreover, these layers do not capture temporal interactions across frames, which are potentially important for robust uncertainty estimation. Second, the uncertainty model in xi-vector is trained implicitly, relying solely on classification loss without explicit supervision of the uncertainty estimates.

To address these issues, we propose the xi+, which enhances uncertainty modeling through explicit supervision and temporal context integration. The main contributions of this paper are summarized as follows:
\begin{enumerate}
\item We introduce a Transformer-based uncertainty estimation module that enhances  model capacity and captures temporal dependencies across frames, improving the expressiveness of the precision matrix prediction.
\item We propose a novel loss function, the Stochastic Variance Loss (SVL), which explicitly supervises the  uncertainty model training. 
\item We incorporate the learned uncertainty into the cosine scoring back-end via a revised uncertainty-aware similarity metric, further improving recognition performance under variable conditions.
\end{enumerate}



\vspace{-2mm}
\section{NEURAL SPEAKER EMBEDDINGS}
\vspace{-2mm}



\subsection{Recap of Xi-vector embedding with Uncertainty}
\vspace{-1mm}

\begin{figure}[t]
\vspace{-3mm}
    \centering
    \includegraphics[width=0.8\linewidth]{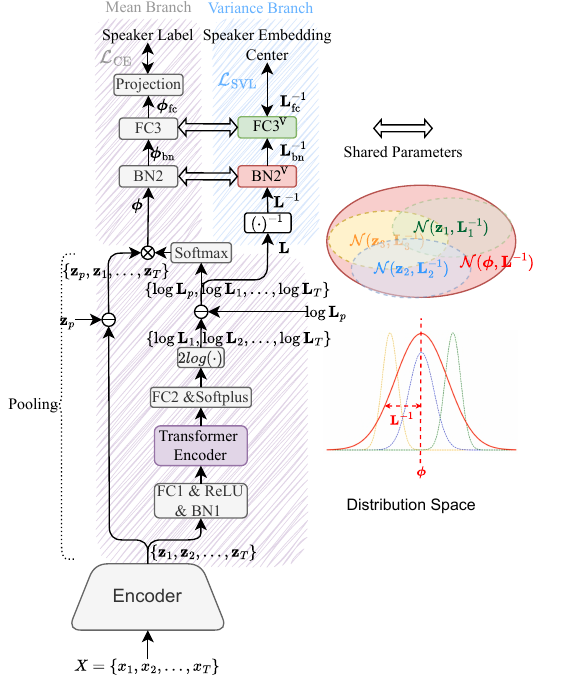}
    \vspace{-3mm}
    \caption{ The architecture of xi+, an extended version of the xi-vector model \cite{lee2021xi}. The modules shown in gray correspond to components of the original xi-vector system, while the modules highlighted in other colors represent our proposed extensions. In the diagram, $\mathbf{z}_p$ denotes the prior mean vector, and $\mathbf{L}_p$ denotes the prior precision matrix. $\otimes$ and $\ominus$ denotes element-wise multiplication and concatenation, respectively. BN$^\text{V}$ and FC3$^\text{V}$ constitute the variance-processing branch, which is structurally parallel to the mean branch. Specifically, they share parameters with BN2 and FC3 in the mean branch, ensuring consistency across the two branches. As shown in distribution space,  all frame-level Gaussian distributions are integrated to form an utterance-level Gaussian distribution, which serves as a compact representation of the speaker. }
    \vspace{-3mm}
    \label{fig:model}
\end{figure}

The xi-vector architecture, proposed in \cite{lee2021xi}, incorporates a frame-level uncertainty estimation mechanism within the pooling layer, as illustrated in Fig. \ref{fig:model}.  The modules in gray represent the original xi-vector system, while the components shown in other colors correspond to the contributions introduced in this paper, which will be introduced in subsequent sections. 

Given a speech utterance $X=\{x_1, x_2, ... , x_T\}$ with $T$ frames, the $t$-th frame is mapped to a latent representation $\mathbf{z}_t$ by the encoder.  In addition, the corresponding precision matrix $\mathbf{L}_t$, which captures the uncertainty associated with the point estimate $\mathbf{z}_t$, is also estimated. To reduce computational complexity, $\mathbf{L}_t$ is constrained to be a diagonal matrix. 

From the perspective of a linear Gaussian model \cite{bishop2006pattern}, the observed representation $\mathbf{z}_t$
  is assumed to consist of two components:  a latent variable $\mathbf{h}$ representing the underlying speaker identity and a random variable $\boldsymbol{\epsilon}_t \sim \mathcal{N}(0,\mathbf{L}_t^{-1})$ catering for the uncertainty for each frame: 
\vspace{-1mm}
\begin{equation}
    \mathbf{z}_t = \mathbf{h}+ \boldsymbol{\epsilon}_t.
\end{equation}

In addition, xi-vector assumes the prior mean  and prior covariance of $\mathbf{h}$ as $\mathbf{z}_p$ and $\mathbf{L}_p^{-1}$\footnote{The prior mean  $\mathbf{z}_p$
  and the prior covariance $\mathbf{L}_p^{-1}$
  are learnable parameters and initialized as $0$ and $\mathbf{I}$, respectively, implying that the prior distribution of the latent variable $\mathbf{h}$
 is assumed to follow a standard normal distribution.}. Mathematically, the posterior distribution of $\mathbf{h}$ is estimated as follows: 
 \vspace{-1mm}
\begin{equation}
p(\mathbf{h}|\mathbf{z}_1,...,\mathbf{z}_T,\mathbf{L}_1^{-1},...,\mathbf{L}_T^{-1})=\mathcal{N}(\mathbf{h}|\boldsymbol{\phi},\mathbf{L}^{-1}),
\end{equation}
where $\mathcal{N}(\cdot)$ denotes gaussian distribution, and the posterior mean vector and precision matrix are computed as follows:
\begin{align}
    \boldsymbol{\phi} &= \frac{\sum_{t=1}^{T}\mathbf{L}_t\mathbf{z}_t+\mathbf{L}_p\mathbf{z}_p}{\mathbf{L}}, \\
    \mathbf{L} &= \sum_{t=1}^T\mathbf{L}_t + \mathbf{L}_p. 
    \vspace{-6mm}
\end{align}

    

    

\subsection{Introducing Xi+}
\vspace{-1mm}

Although the xi-vector architecture effectively incorporates uncertainty information into each frame-level speaker representation, the uncertainty modeling network itself is relatively simplistic, consisting only of a few linear layers followed by non-linear activation functions. As a result, it may be insufficient to capture complex and informative uncertainty patterns. To address this limitation, we propose an enhanced model xi+  built upon the xi-vector framework, as shown in Fig. \ref{fig:model}, with the following key improvements:

\subsubsection{Temporal Modeling}
The original xi-vector architecture predicts the precision matrix using fully connected layers, which only capture relationships along the channel dimension. We hypothesize that temporal dependencies are also informative for uncertainty modeling. To this end, we introduce a Transformer encoder containing a single layer with $N$ attention heads within the pooling layer \footnote{$N$ is set to 8 in this experiment.}. This newly added module serves two primary purposes:  enhancing the non-linear modeling capacity of the network, and  enabling the incorporation of temporal contextual information across frames.

\subsubsection{Stochastic Variance Loss}We hypothesize that the implicit supervision of uncertainty model in xi-vector, which relies solely on the cross-entropy loss $\mathcal{L}_\text{CE}$, is insufficient for accurately capturing frame-level uncertainty. To address this limitation, we propose a novel loss function, the Stochastic Variance Loss $\mathcal{L}_\text{SVL}$, which is 
directly applied on uncertainty estimation branch, explicitly encouraging the model to learn accurate  and well-calibrated uncertainty estimation.  

The key idea of $\mathcal{L}_\text{SVL}$ is to establish a pseudo ground truth for uncertainty. Specifically, we regard the distance between a speaker embedding $\boldsymbol{\phi}_\text{fc}$ and the centroid of embeddings from the same speaker as a pseudo-label. Since each variance estimate is supervised only by a single embedding–centroid pair, rather than the full distribution, we term this approach `stochastic variance' learning.


Usually, each speaker has more than one utterance. The $\mathcal{L}_\text{SVL}$ loss utilizes the centroid of speaker embeddings pertaining to each speaker as target. Assume that the speaker identity for utterance $j$ is $y_j$, then its centroid should be: 
\begin{equation}
    \mathbf{c}^{y_j} = \frac{\sum_{j=1}^{J}  \boldsymbol{\phi}_\text{fc}^{j}}{J},
\end{equation}
where all $J$ utterances belong to the same speaker identity $y_j$. 
  Then the $\mathcal{L}_\text{SVL}$ loss is given by: 

\begin{equation}
    \mathcal{L}_\text{SVL} = \frac{1}{B}  \sum_{b=1}^{B} \|\alpha\sqrt{{(\mathbf{L}_\text{fc}^{b}})^{-1}} -|\boldsymbol{\phi}_\text{fc}^b-\mathbf{c}^{y_b}| \|^2_2,
    \label{equ:new_svl}
\end{equation}
where $B$ denotes the number of batch size in training. $(\mathbf{L}_\text{fc}^{b})^{-1}$ and $\mathbf{c}^{y_b}$
denote the covariance matrix  for utterance $b$ and the centroid of speaker $y_b$, respectively. $\|\cdot\|_2$ denotes  the 
Euclidean norm. In addition, we introduce a learnable scaling parameter $\alpha$, referred to as the uncertainty scaling factor, to address the potential scale mismatch between the predicted standard deviation $\sqrt{(\mathbf{L}_\text{fc}^{b})^{-1}}$ and the observed embedding deviation $|\boldsymbol{\phi}_\text{fc}^{b}-\mathbf{c}^{y_b}|$. 

To propagate the covariance matrix $\mathbf{L}^{-1}$  through the same layers of mean vector $\boldsymbol{\phi}$. Referring to Wang et al. \cite{wang2023incorporating}, 
we design two coordinated branches to jointly model them.  Rather than treating them independently, these branches are constructed to preserve the mathematical relationships between the mean and variance inherent to Gaussian modeling.

For the mean branch, the utterance-level representation is first normalized via batch normalization:
\begin{equation}
    \boldsymbol{\phi}_\text{bn} = \frac{\boldsymbol{\phi} - \boldsymbol{\mu}_\text{bn}}{\sqrt{\boldsymbol{\sigma}_\text{bn}+\boldsymbol{\epsilon}}} \otimes \boldsymbol{\gamma} +\boldsymbol{\beta},
\end{equation}
where $\boldsymbol{\mu}_\text{bn}$ and $\boldsymbol{\sigma}_\text{bn}$ denote the batch-wise mean and variance, respectively.  
$\boldsymbol{\epsilon}$ is a small constant for numerical stability, and $\boldsymbol{\gamma}$ and $\boldsymbol{\beta}$ are  learnable affine parameters. The symbol $\otimes$ denotes element-wise multiplication. 
Subsequently, the normalized vector is passed through a fully connected layer:
\begin{equation}
    \boldsymbol{\phi}_\text{fc} = \boldsymbol{\phi}_\text{bn} \mathbf{W}^\top + \mathbf{b}, 
\end{equation}
where $\mathbf{W}$ and $\mathbf{b}$ are   learnable parameters in FC3 in Fig. \ref{fig:model}. 

The variance branch shares its parameters with the mean branch, but adopts a distinct computational formulation to model the covariance structure. 
For the batch normalization step, the input precision matrix is transformed as follows:
\begin{equation}
    \mathbf{L}_\text{bn}^{-1} = \frac{\mathbf{L}^{-1} \otimes \mathbf{\boldsymbol{\gamma}}^2}{\boldsymbol{\sigma}_\text{bn}+\boldsymbol{\epsilon}}.
\end{equation}
Subsequently, the normalized vector is passed through a fully connected layer: 
\begin{equation}
    \mathbf{L}_\text{fc}^{-1} = \mathbf{W}\mathbf{L}_\text{bn}^{-1}\mathbf{W}^\top. 
\end{equation}

\subsubsection{Cosine Scoring with Uncertainty}

We propose to incorporate uncertainty information into the cosine scoring process, as shown below: 
\begin{equation}
    s_{u} = \frac{<\boldsymbol{\phi}^{s2}_\text{fc},\boldsymbol{\phi}^{s1}_\text{fc}>}{\sqrt{\boldsymbol{\phi}{^{s2}_\text{fc}}^\top (\mathbf{I}+\rho(\mathbf{L}_\text{fc}^{s2})^{-1})^{-1}\boldsymbol{\phi}^{s2}_\text{fc}}  \sqrt{\boldsymbol{{\phi}}{^{s1}_\text{fc}}^\top (\mathbf{I}+\rho(\mathbf{L}^{s1}_\text{fc})^{-1})^{-1}\boldsymbol{\phi}^{s1}_\text{fc}}}.
    \label{equ:cos}
\end{equation}
Here, $\rho$ denotes a scaling factor in cosine scoring, and $\mathbf{I}$ is the identity matrix. The superscripts $s1$ and $s2$ indicate the corresponding speaker identities for the enrollment and test embeddings, respectively.  When the covariance matrix or $\rho$ is zero, the uncertainty-aware similarity score $s_u$ reduces to the normal cosine similarity.

In our previous work \cite{wang2024cosine}, $\rho$ is set to a fixed number,  $\frac{1}{d}$, where $d$ is the dimension of the speaker embedding.  However, we argue that the fixed scaling factor $\frac{1}{d}$ may not be optimal. Instead, this paper proposes to adopt the learnable uncertainty scaling factor $\alpha$, derived from the proposed loss $\mathcal{L}_\text{SVL}$, as a more effective alternative. Unlike the static $\frac{1}{d}$, $\alpha$ is learned in a data-driven manner to reflect the actual uncertainty scale observed in the embedding space, thus we believe it could provide a more adaptive and accurate scaling of the covariance term in the scoring function.

\vspace{-1mm}

\section{Experiments}
\vspace{-2mm}

\subsection{Training details}
\vspace{-1mm}
We adopt the stage 1 to stage 5 training pipeline from the VoxCeleb v2 recipe provided by the WeSpeaker toolkit \cite{wang2023wespeaker,wang2024advancing}\footnote{\url{https://github.com/wenet-e2e/wespeaker/blob/master/examples/voxceleb/v2/run.sh}}, following all default hyperparameter settings. The training spans 150 epochs, using 2-second audio segments. AAM-Softmax \cite{deng2019arcface} is applied with a scaling factor of 32. The angular margin is initially set to zero and then progressively increased from 0 to 0.2 between epochs 20 and 40, after which it remains fixed. The learning rate schedule begins with a linear warm-up over the first 6 epochs, rising from 0 to a peak of 0.1, followed by an exponential decay down to 5e-5 for the remaining training process. Finally, we average the parameters from the last 10 checkpoints to obtain a final model checkpoint. The dimension of speaker embedding $d$ is 192. 

We also perform large-margin fine-tuning, referred to as `LM', using the WeSpeaker framework for some models. During this stage, the total number of training epochs is set to 5, with an angular margin of 0.5, and 9-second audio segments are used, following the findings of Barahona et al. \cite{barahona2025analysis}, who suggest that longer segments can lead to consistent improvements. The learning rate gradually decreases from $1 \times 10^{-4}$   to $2.5 \times 10^{-5}$. We adopt the last checkpoint as a final model checkpoint in this stage. 

To obtain the centroid embedding for each speaker, we first fully pre-train a model using the same architecture as xi+, but without
$\mathcal{L}_\text{SVL}$. This pre-trained model is then used to extract speaker embeddings from each speaker's utterance in training set and compute the centroid. Afterward, we apply the 
$\mathcal{L}_\text{SVL}$ 
  as follows:
\begin{equation}
\mathcal{L}_{\text{Final}} = \mathcal{L}_{\text{CE}}+ \kappa \mathcal{L}_\text{SVL},
\label{equ:loss}
\end{equation}
where $\kappa$  is used to control loss weight and  is dynamically adjusted based on the current epoch $ep$ to  ensure that $\mathcal{L}_\text{SVL}$ 
  gradually contributes as training progresses: 
\begin{equation}
  \kappa = 
  \begin{cases}
    0, & \text{if } \text{epoch} \leq ep_\text{SVL}, \\
   \lambda \frac{ep-ep_\text{SVL}}{ep_\text{Max}-ep_\text{SVL}}, & \text{otherwise},
  \end{cases}
\end{equation}
where $\lambda$ is a pre-defined fixed constant, $ep_\text{SVL}$ indicates epoch when $\mathcal{L}_\text{SVL}$ starts to be applied. $ep_\text{Max}$ indicates the maximum number of training epochs.  In our experiments, we set $ep_\text{SVL}=70$ \footnote{We think if $ep_\text{SVL}$ is too small, the speaker encoder may not have learned sufficiently robust representations, which could negatively impact the final performance.}    and $ep_\text{Max}=150$.

\renewcommand{\arraystretch}{1.3}
\begin{table*}[htbp]
\centering
 \caption{Overall results on Voxceleb1 in terms of the
EER (\%) and MinDCF (\textit{P}target = 0.01). $\lambda$ denotes the loss weight in Equation \ref{equ:loss}, while $\alpha$ denotes the uncertainty scaling factor in $\mathcal{L}_\text{SVL}$.  $\rho$ denotes the scaling factor used in cosine scoring.  The gray color highlights our best-performing configuration. All experiments in this table employ ECAPA-TDNN \cite{desplanques20_interspeech} as the speaker encoder. }
\begin{tabular}{l|l|l|l|l|l|l|ll|ll|ll}
\hline
\multirow{2}{*}{\#Exp. } & \multirow{2}{*}{Model} & \multirow{2}{*}{\#Param.}    & Temporal       & \multirow{2}{*}{ $\lambda$}   & \multirow{2}{*}{ $\alpha$}           & \multirow{2}{*}{$\rho$}        & \multicolumn{2}{l|}{Vox1-O}         & \multicolumn{2}{l|}{Vox1-E}         & \multicolumn{2}{l}{Vox1-H}         \\ \cline{8-13} 
                    &   &                           &                                            Modeling         &                                &    &                              & \multicolumn{1}{l|}{EER}   & minDCF & \multicolumn{1}{l|}{EER}   & minDCF & \multicolumn{1}{l|}{EER}   & minDCF \\ \hline
1 \footnotemark & x-vector \cite{snyder2018x,desplanques20_interspeech}             & 6.19M                      & \multirow{2}{*}{\ding{56} }                 & \multirow{3}{*}{$0$}     & \multirow{4}{*}{\ding{56} }           & \multirow{5}{*}{$0$}      & \multicolumn{1}{l|}{1.069} &  -     & \multicolumn{1}{l|}{1.209} &    -    & \multicolumn{1}{l|}{2.310} &    -    \\ \cline{1-3} \cline{8-13} 
2& xi-vector \cite{lee2021xi}              & 5.90M                                                 &     &                                &                                            &                                 & \multicolumn{1}{l|}{0.995} & 0.103  & \multicolumn{1}{l|}{1.130} & 0.126  & \multicolumn{1}{l|}{2.169} & 0.209  \\ \cline{1-3} \cline{4-4} \cline{8-13} 
3& \multirow{7}{*}{xi+}   & \multirow{7}{*}{6.69M}                                 & \multirow{7}{*}{\ding{52}}                 &                                            &         &                        & \multicolumn{1}{l|}{0.995} & 0.089  & \multicolumn{1}{l|}{1.127} & 0.124  & \multicolumn{1}{l|}{2.106} & 0.206  \\ \cline{1-1}\cline{5-5} \cline{8-13} 
  4&                     &                           &                                                                 &\multirow{4}{*}{$0.01$} &              &                                 & \multicolumn{1}{l|}{1.031} & 0.093  & \multicolumn{1}{l|}{1.120} & 0.125  & \multicolumn{1}{l|}{2.148} & 0.210  \\ \cline{1-1} \cline{6-6}  \cline{8-13} 
                    5&   &                           &                             &                                    &  &                                   & \multicolumn{1}{l|}{0.984} & 0.092  & \multicolumn{1}{l|}{1.118} & 0.123  & \multicolumn{1}{l|}{2.122} & 0.210  \\\cline{1-1} \cline{7-7} \cline{8-13} 
                      6 &   &                        &                             &                                    &                                            & $1/d$                     & \multicolumn{1}{l|}{0.973} & 0.087  & \multicolumn{1}{l|}{1.106} & 0.121  & \multicolumn{1}{l|}{2.110} & 0.209  \\ \hhline{|-~~~~~-------|} 
                       \rowcolor{gray!20}7&  & &   &                                  &         \textcolor{black}{\ding{52}}   &  \multirow{3}{*}{$\alpha$} & \multicolumn{1}{l|}{ \textbf{0.899}} &  0.093  & \multicolumn{1}{l|}{ \textbf{1.043}} &  \textbf{0.115}  & \multicolumn{1}{l|}{ 2.071} &  0.213  \\ \cline{1-1}\cline{5-5}\cline{8-13} 
                    8&   &                                                        &                                    & $0.05$   &               &                                 & \multicolumn{1}{l|}{0.936} & 0.084  & \multicolumn{1}{l|}{1.047} & 0.116  & \multicolumn{1}{l|}{\textbf{2.063}} & 0.210  \\ \cline{1-1}\cline{5-5} \cline{8-13} 
                    9 &    &                                                       &                                    & $0.1$  &                   &                                 & \multicolumn{1}{l|}{0.962} & \textbf{0.081}  & \multicolumn{1}{l|}{1.065} & 0.116  & \multicolumn{1}{l|}{2.064} & \textbf{0.204}  \\ \hline



\end{tabular}
\label{tab:result1}
\vspace{-5mm}
\end{table*}
\footnotetext{Results come from \url{https://github.com/wenet-e2e/wespeaker/blob/master/examples/voxceleb/v2/README.md}}

\subsection{Dataset}
\vspace{-1mm}

We conduct two sets of experiments. The first set is performed on Voxceleb2, which consists of 5,994 speakers, and is evaluated on Voxceleb1 \cite{nagrani2020voxceleb}. The second set is conducted on the NIST SRE 2024 dataset. We follow the evaluation plan outlined in \cite{nist2024} to construct our training, development, and evaluation datasets. Additionally, we apply rVAD \cite{tan2020rvad} to remove non-speech signals. For both sets, we also employ data augmentation techniques, incorporating noise from the MUSAN database \cite{snyder2015musan} and room impulse responses from the RIR database \cite{ko2017study}.

\vspace{-2mm}
\section{Results}
\vspace{-2mm}


\subsection{Results on Voxceleb}
\vspace{-1mm}

Table \ref{tab:result1} presents the overall results on VoxCeleb1. The x-vector using channel- and context-dependent statistics pooling \cite{desplanques20_interspeech}, does not incorporate any uncertainty estimation. In contrast, the xi-vector model includes an implicit uncertainty estimation mechanism, while the proposed xi+ model with $\mathcal{L}_\text{SVL}$ introduces an explicit uncertainty modeling strategy. 

Exp 1 and Exp 2 demonstrate that incorporating uncertainty estimation is beneficial for speaker embedding, with Exp 2 achieving superior performance. In Exp 3, a Transformer encoder is introduced to model temporal dependencies across frames, leading to consistent improvements across all evaluation metrics. This demonstrates the effectiveness of incorporating temporal modeling for capturing frame-level uncertainty.  Exp 4 and 5 further incorporate explicit uncertainty supervision using the proposed loss $\mathcal{L}_\text{SVL}$ ($\lambda>0$). Results show that Exp 5, which includes a learnable scaling parameter $\alpha$  in $\mathcal{L}_\text{SVL}$, achieves overall performance improvements, although performance on the Vox1-H set is slightly lower than that of Exp 3.

Notably, Exp 6 and 7 apply uncertainty-aware cosine scoring, leading to significant performance gains. Among them, Exp 7 which uses the proposed learnable uncertainty scaling factor 
$\alpha$ \footnote{The learned value of $\alpha$ in Exp 7 is 0.0573 after training.} obtained from $\mathcal{L}_\text{SVL}$ in the cosine scoring function, achieving particularly strong results. This demonstrates that the proposed $\mathcal{L}_\text{SVL}$
  effectively learns a meaningful uncertainty scaling factor, which can be beneficial when integrated into the cosine scoring function. In addition, we investigate the impact of the loss weight $\lambda$ by varying its value in Exp 8 and 9.
Overall, Exp 7 achieves the best performance across most evaluation conditions, yielding 9.6\%, 7.7\%, and 4.5\% relative improvements on Vox1-O, Vox1-E, and Vox1-H, respectively, compared to the xi-vector in terms of EER.  In the following section, we adopt this configuration as the default setup for further analysis and evaluation.



\renewcommand{\arraystretch}{1.3}
\begin{table}[htbp]
\centering
 \caption{Comparison of different systems employing cosine scoring on on SRE 2024 development and evaluation sets in terms of EER. xi+ adopts the best configuration from  Table \ref{tab:result1}, which is highlighted in gray. `LM' denotes large margin fine-tuning \cite{thienpondt2021idlab}. In this table, a new speaker encoder Golden Gemini \cite{liu2024golden} is adopted for all models.}
\begin{tabular}{l|l|l|l|l|l}
\hline
\multirow{2}{*}{Model } & \multirow{2}{*}{\#Param.} & \multirow{2}{*}{LM} & $\rho$ &  \multicolumn{2}{c}{EER}  \\ \cline{5-6}
& & & &    dev & eval  \\ \hline
\multirow{2}{*}{x-vector \cite{snyder2018x, liu2024golden}} & \multirow{2}{*}{9.04M} &     \ding{56} & \multirow{4}{*}{0}  & 17.450& 18.381\\ \cline{3-3} \cline{5-6}
& &  \ding{52}  & & 13.909 & 14.280\\ \cline{1-3} \cline{5-6}
\multirow{2}{*}{xi-vector \cite{lee2021xi}} & \multirow{2}{*}{9.49M} &     \ding{56} &   &16.298 & 16.842 \\ \cline{3-3} \cline{5-6}
& &  \ding{52} &  & 13.265 &13.342 \\ \cline{1-2} \cline{3-6}
 & &  \multirow{3}{*}{\ding{56}} & 0 &  16.602 & 16.018\\  \cline{4-6}
& &  & $1/d$ & 16.570 & 15.996\\  \cline{4-6} \hhline{|~~~---|} 
 \rowcolor{gray!20} xi+&  10.24M & &  $\alpha$ &16.196  & 15.626 \\ \cline{3-6}
& &  \multirow{3}{*}{\ding{52}} & 0& 13.024 & 12.233\\\cline{4-6}
& &   & $1/d$& 12.928&12.149 \\ \cline{4-6} \hhline{|~~~---|} 
\rowcolor{gray!20} & &  & $\alpha$ & \textbf{12.265} & \textbf{11.745} \\ \hline

\end{tabular}

\label{tab:result2}
\vspace{-3mm}
\end{table}
\vspace{-1mm}
\subsection{Results on SRE}
\vspace{-1mm}
Table~\ref{tab:result2} presents the performance of different models on the NIST SRE 2024 dataset. Since SRE data is more complex and challenging to model, we adopt a more powerful speaker encoder, Golden Gemini \cite{liu2024golden}. This also allows us to evaluate the generalization capability of our proposed methods across different encoder architectures. 
In this setting, we apply a large-margin fine-tuning (LM) strategy, which significantly improves model performance, yielding a relative gain of approximately 20\% in terms of EER (e.g., reducing the metric from 16.196\% to 12.2651\% on the dev set). Notably, our proposed model, xi+, demonstrates consistent improvements over the baseline: without LM, the score decreases from 16.298\% to 16.196\% on dev set and from 16.842\% to 15.626\% on eval set; with LM, it further drops to 12.265\%  on dev set and  11.745\% on eval set, achieving a relative improvement of  8\% on dev set and 11\% on eval set \footnote{The learned value of $\alpha$ is 0.1576 without LM, and 0.1849 with LM fine-tuning.}.

\vspace{-2mm}

\section{Conclusions}
\vspace{-2mm}
In this paper, we propose several enhancements to address the limitations of the xi-vector framework. First, we introduce a temporal modeling mechanism to capture the dynamics of uncertainty across temporal dimension. Second, we propose a novel loss function, $\mathcal{L}_\text{SVL}$, to explicitly supervise the learning of uncertainty estimation, which is the key contribution in this paper.  Third, we incorporate a learnable uncertainty scaling factor into the cosine scoring process to better reflect confidence in speaker embeddings. We conduct ablation studies with controlled variables to systematically analyze and verify the effectiveness and individual contribution of each component, with more pronounced improvements observed on the more challenging SRE dataset.

\vspace{-2mm}
\section{ACKNOWLEDGMENTS}
\vspace{-1mm}
This work was supported in part by computing resources provided by the University Research Facility in Big Data Analytics (UBDA) at The Hong Kong Polytechnic University, and in part by the General Research Fund (GRF) of the Research Grants Council of the Hong Kong SAR under Project No. 15228223.

\bibliographystyle{IEEEbib}
\bibliography{refs}

\end{document}